\documentclass[journal]{IEEEtran}
\usepackage{amsmath,amssymb,amsfonts}
\usepackage{algorithm}
\usepackage{algorithmic}
\usepackage{graphicx}
\usepackage{textcomp}
\usepackage{cite}
\usepackage{mathtools}
\usepackage{bm}
\usepackage{xcolor}
\usepackage{booktabs}
\usepackage{array}
\usepackage{multirow}
\usepackage{needspace}
\usepackage{url}
\usepackage[bookmarks=false,hidelinks]{hyperref}
\hypersetup{
  pdftitle={Radar-Only Long-Term Beam Prediction in mmWave Vehicular Networks: An Empirical Study of Doppler Input and Cross-Modal Distillation},
  pdfauthor={Jeongwan Kang, Miguel Bordallo L\'{o}pez, Nhan Thanh Nguyen, Junhyeong Kim, and Sunwoo Kim}
}
\usepackage{pifont}
\usepackage{ifthen}

\DeclareMathOperator*{\argmax}{arg\,max}

\begin{document}
\bstctlcite{IEEEexample:BSTcontrol}

\title{Radar-Only Long-Term Beam Prediction in mmWave Vehicular Networks:
An Empirical Study of Doppler Input and Cross-Modal Distillation}

\author{Jeongwan Kang,~\IEEEmembership{Member,~IEEE}, Miguel Bordallo L\'{o}pez,~\IEEEmembership{Member,~IEEE}, \\ Nhan Thanh Nguyen,~\IEEEmembership{Senior Member,~IEEE}, Junhyeong Kim, and Sunwoo Kim,~\IEEEmembership{Senior Member,~IEEE}
\thanks{This work was supported by the Institute of Information \& Communications Technology Planning \& Evaluation (IITP) grant funded by the Korea government (MSIT) (No. RS-2024-00444230, Development of wireless technology for integrated sensing and communication), and supported in part by Business Finland, Nokia, Keysight and Finwe via the 6G-WISECOM project.
 J. Kang and S. Kim are with the Department of Electronic Engineering, Hanyang University, Seoul, Republic of Korea (e-mail: \{rkdwjddhks77, remero\}@hanyang.ac.kr). M. Bordallo L\'{o}pez is with Computer Science and Engineering, and N. T. Nguyen is with the Centre for Wireless Communications, University of Oulu, Oulu, Finland (e-mail: \{miguel.bordallo, nhan.nguyen\}@oulu.fi). J. Kim is with the 6G Wireless Access System Research Section, Electronics and Telecommunications Research Institute (ETRI), Daejeon, Republic of Korea. He is also with the Major in Information and Communication Engineering, ETRI School, University of Science and Technology (UST), Daejeon, Republic of Korea (e-mail: jhkim41jf@etri.re.kr). (Corresponding author: Sunwoo Kim.)}}

\maketitle

\begin{abstract}\boldmath
Sensing-aided beam prediction could cut the beam-training overhead of millimeter-wave vehicular links,
and two design choices could strengthen a radar-only predictor: a Doppler-resolved input and
knowledge distilled from a camera during training. We measure what each is worth in received
power rather than in top-$1$ accuracy, whose differences of a few points amount to only tenths
of a decibel. We use real DeepSense~6G measurements, mainly at one roadside site, and one
convolutional--recurrent predictor family evaluated open-loop up to $0.67$\,s ahead. Adding a
range--Doppler map to the range--angle map lowers the mean beam-selection loss
by $0.28$\,dB ($95\%$ bootstrap interval $[0.14, 0.41]$). The gain comes from the Doppler
content, not from the extra input, and replicates, as pre-specified, at a second base station,
although holding the last measured beam beats both predictors there. Cross-modal knowledge distillation
from a motion-mask vision--radar teacher, by contrast, changes received power by $-0.01$\,dB
($[-0.07, +0.06]$). The tested teacher, student, fusion and distillation
variants show the same null in the metric each was scored on. Where distillation helps, in an exploratory run with few
training passes, a radar-only teacher trained on all training passes reproduces most of the gain. For this predictor family and the full training set, the measurable gain at the link comes from
Doppler-resolved radar content; no tested route for transferring training-only visual
information adds a resolved gain.
\end{abstract}


\section{Introduction}
\label{sec:intro}
\IEEEPARstart{M}{illimeter-wave} (mmWave) communication with large-scale antenna arrays
is a cornerstone of next-generation vehicular networks~\cite{niu2015survey, jiang20216g}.
To overcome the severe path loss, the base station~(BS) concentrates energy into narrow
beams steered toward the vehicle~\cite{heath2016overview}. Under mobility, that alignment
must be updated continuously: a stale beam lowers the received signal-to-noise ratio and
can drop the link. Conventional beam training scans a codebook to find the strongest beam,
but its overhead grows with the codebook size and becomes prohibitive under
mobility~\cite{yi2024beam}. Sensing-aided beam management, closely related to integrated
sensing and communication~\cite{cheng2022isac}, offers an alternative: sensors co-located
with the BS let a learned model predict the beam from what they observe, without an
exhaustive over-the-air sweep.

Sensing-aided predictors are often heavy and depend on rich sensor suites, at odds with
low-cost roadside infrastructure. Two choices suit roadside deployment but make
prediction harder.
\emph{Long-term} prediction, which forecasts several future slots from one observation
window, amortizes sensing and inference over the horizon, but the predictor must
anticipate where the beam will move. Radar measures radial velocity directly, a natural
cue to that motion. Intuitively, a range--angle map tells a predictor where the vehicle is, and
only a sequence of such maps reveals where it is heading. A Doppler-resolved map adds, within
a single frame, a measure of how fast each moving return approaches or recedes. On a road of known
geometry, position and radial velocity fix the velocity vector away from broadside, and hence the
vehicle's next angle, an intuition Sec.~\ref{sec:res_doppler} tests. A \emph{radar-only} predictor raises fewer privacy concerns and does
not depend on illumination, radar being an active sensor, but it loses what the camera
sees. Cross-modal knowledge distillation (KD)~\cite{nguyen2026kd}, in which a vision--radar
teacher trains a lightweight radar-only student, is one candidate way to recover that
information. The teacher sees the vehicle in the camera image during training and can steer the
student toward the beams that image supports, while the student still reads radar alone at
inference. Distillation can help only to the extent that the teacher's advantage is information the
radar maps can also express.

Three gaps remain (Sec.~\ref{sec:related}); the first concerns the scale needed to judge the
other two. \emph{Gap~1 (accuracy without a decibel scale):} this literature ranks designs by
top-$k$ accuracy, which scores a neighboring beam as wrong as a distant one. Link-level effect sizes are rarely
reported, so whether a one- or
two-point difference matters in received power is unknown. \emph{Gap~2 (Doppler only as
a single-slot replacement):} predictors whose only input is radar have used the Doppler
dimension only in place of the range--angle map, and only for the current slot. In that
setting, range--velocity maps and full radar cubes were no better than range--angle
input~\cite{demirhan2022radar}. Whether a Doppler-resolved input helps a radar-only
predictor free of channel state information (CSI) to forecast several slots at once, and, if
so, what information it must preserve, has not been measured. \emph{Gap~3 (an untested
transfer premise):} the reported benefit of cross-modal KD rests on \emph{simulated} data,
\emph{single-step} prediction and a radar student fed a simulated point cloud rather than
measured Doppler maps~\cite{park2025ckd}. KD for long-term prediction has been shown only for \emph{same-modality}
transfer~\cite{ma2025kd}. It has not been tested whether that benefit survives on real
measurements, over a long horizon and for a student that already reads a Doppler-resolved
map.

We therefore ask how much received power a Doppler-resolved input and a training-only
camera add to a radar-only long-term beam predictor. Every prediction is scored by its
loss in received power against the best beam of the recorded sweep (Gap~1). A
Doppler-resolved radar-only predictor is the testbed for Gap~2, and a cross-modal
distillation pipeline for Gap~3; the contribution is the controlled comparison, including its
null result, not the architecture. All experiments use the real-world DeepSense~6G
dataset~\cite{10144504}, mainly at a straight-road site where one transmitter vehicle passes in
one direction at typical speeds of $2.8$--$4.7$\,m/s. The high-mobility regime that motivates long-term
prediction is therefore argued, not tested.

\subsection{Contributions}
\label{sec:contrib}
We make three contributions, positioned against prior work in Sec.~\ref{sec:rel_position};
C1 provides the scale for C2 and C3:

\begin{itemize}
  \item \textbf{(C1) Received power as the common scale.} We measure what top-$1$ accuracy is
  worth in received power, using the beam powers recorded with the labels. Accuracy orders
  the tested predictors almost as received power does, but its typical gaps
  amount to only tenths of a decibel. Every principal contrast in this paper is
  therefore measured in decibels (Sec.~\ref{sec:res_resolution}).

  \item \textbf{(C2) A replicated Doppler gain, with bounds on its source.} We compare radar
  maps with and without a Doppler axis: a
  Doppler-resolved input gives a measurable link-level gain, mostly in the tail of the loss, to the tested
  long-term radar-only predictors. Controls on capacity and input content trace the gain to the
  Doppler content, not to the extra input, and a pre-specified test replicates it at a
  second BS site (Secs.~\ref{sec:res_pred} and~\ref{sec:res_doppler}).

  \item \textbf{(C3) A controlled null result for cross-modal distillation.} On real
  data, we vary the teacher, the student, the fusion and the distillation loss under one protocol.
  With the full training set, distilling a motion-mask vision--radar teacher adds no measurable
  received power in the variants scored in decibels and no measurable accuracy gain in those
  scored only in accuracy. With scarce data (exploratory), distillation does add received power,
  and a radar-only teacher trained on all training passes reproduces most of that gain
  (Secs.~\ref{sec:res_ablation} and~\ref{sec:res_fusion}).

\end{itemize}

Sec.~\ref{sec:related} positions the work; Secs.~\ref{sec:system}--\ref{sec:predictor} give
the measured-link model and the testbed. Sec.~\ref{sec:results} gives the protocol, accuracy
and principal contrasts; Sec.~\ref{sec:res_resolution} the received-power scale (C1);
Sec.~\ref{sec:res_choices} the design choices on that scale; Sec.~\ref{sec:conclusion}
concludes. S-numbered sections, tables and figures are in the supplement.

\emph{Overlap disclosure.} Our companion paper~\cite{companion_voi} uses the same dataset and cached
predictions to ask when invoking a predictor is worthwhile relative to feedback in a
causal closed loop. It reports the feedback-only baselines, causal replay
and the probe-verified invocation controller; this paper reports the
predictor and contributions C1--C3, under the windowed, open-loop protocol. Scenario~9 is the discovery site, its test passes having guided the design
(Sec.~\ref{sec:setup}). Scenario~31 enters as a pre-specified replication of the
Doppler contrast, and Scenarios~32 and~33 only in an inconclusive day/night check (Sec.~\ref{sec:res_pred}). The
results below are self-contained.

\section{Related Work}
\label{sec:related}
Sec.~\ref{sec:rel_longterm} surveys sensing-aided and long-term beam prediction;
Secs.~\ref{sec:rel_eval}--\ref{sec:rel_kd} then locate Gaps~1--3 in turn, and
Sec.~\ref{sec:rel_position} positions this work.

\subsection{Sensing-Aided and Long-Term Beam Prediction}
\label{sec:rel_longterm}
Deep learning models map sensor observations to beam indices, replacing costly over-the-air
training. Single-modality predictors have been demonstrated with
cameras~\cite{9923616, imran2024env}, radar~\cite{demirhan2022radar} and
LiDAR~\cite{klautau2019lidar}. Multimodal methods fuse vision with
position~\cite{charan2022vision} or combine several sensors through
Transformers~\cite{Tian2023MultimodalTF, cui2024sensing} or
attention~\cite{shang2025multimodal, zhu2025crossmodal} for accuracy and robustness
beyond any single sensor. These works predict the current beam with high-capacity
models, leaving both the long-term and the efficiency questions open. Forecasting several
future slots from one observation window has been studied with
LiDAR~\cite{jiang2023lidar}, radar~\cite{luo2023radar} and vision~\cite{jiang2022cv,
ma2025kd, ma2026attention}, using recurrent and attention models. Most forecasters read
one sensing modality; the multimodal ones~\cite{ma2026lightweight,
orimogunje2026beamguard} keep every sensor at inference, so a forecaster that learns
from several modalities but runs on one remains unexplored.

\subsection{How These Methods Are Evaluated}
\label{sec:rel_eval}
These predictors are mostly ranked by top-$k$ accuracy over the beam codebook, often
beside the distance-based accuracy (DBA)~\cite{ma2025kd, charan2022challenge}, which
gives partial credit to predictions within $\Delta$ indices of the optimum. DBA
acknowledges that a near miss may be cheap, but $\Delta$ is a convention, not a
measurement, and an index is not a unit of received power. A one- or two-point
difference therefore need not reflect a meaningful change in received power. A few works report
link-level quantities instead: transmission rate~\cite{9923616} and received signal
strength~\cite{park2025resource} in simulation, or the mean fraction of the best beam's
power that the selected beam retains~\cite{orimogunje2026beamguard}. None asks what a point
of accuracy is worth in received power. The sweeps that produce
the labels also record the received power of every beam, and
Sec.~\ref{sec:res_resolution} uses these powers to answer that question (Gap~1).

\subsection{Radar Representations}
\label{sec:rel_radar}
Where radar is the only input, the Doppler (radial-velocity) dimension has been tried
only in place of the range--angle map and only for the current slot. Demirhan and
Alkhateeb~\cite{demirhan2022radar} compare three radar-only representations for
\emph{current-slot} prediction (range--angle, range--velocity and the full radar cube)
and find the range--angle map the \emph{best} of the three. The range--velocity
(range--Doppler) map is close behind, a closeness they attribute to the radial velocity
carrying angular information on their one-directional road. The full cube, which carries
strictly more information, is no better either. Doppler maps do enter beam
forecasting, in a vehicle-to-vehicle tracker that also reads the previously confirmed
beam~\cite{luo2023radar} and in a multimodal tracker that keeps the
camera~\cite{ma2026lightweight}. Three questions have therefore not been measured:
whether a Doppler-resolved map helps a radar-only, CSI-free predictor over a long
horizon, which projection should carry Doppler and whether the range--angle map is still
needed beside it. Sec.~\ref{sec:res_pred} measures all three (Gap~2).

A separate line, radar-assisted predictive beamforming~\cite{liu2020radar,
yuan2021bayesian} and Kalman-type beam tracking~\cite{atiekf2024}, predicts the beam from a
tracked vehicle state. It needs a detection front end and, in~\cite{liu2020radar,
yuan2021bayesian}, echoes of the BS's own communication signal, whereas the learned
predictor reads the radar maps directly, without detection or tracking.
Sec.~\ref{sec:res_trivial} builds a reduced form of that line as a baseline: an extended
Kalman filter in road coordinates that reads range, angle and Doppler from the same maps.

\subsection{Knowledge Distillation for Efficient Beam Prediction}
\label{sec:rel_kd}
Knowledge distillation trains a lightweight student to reproduce a high-capacity teacher.
It has been applied across wireless tasks~\cite{nguyen2026kd}, including transceiver
design~\cite{kong2021kd}, channel feedback~\cite{tang2021kd, guo2022environment},
semantic communication~\cite{liu2023kd}, positioning~\cite{alahmadi2024kd} and, most
recently, the mapping of sub-6\,GHz channels to mmWave beams~\cite{tavakolian2026sub6}. For
beam prediction, three settings are most relevant. Ma~et~al.~\cite{ma2025kd}
distill a vision teacher into a vision student for long-term prediction, i.e.,
\emph{same-modality} transfer. The same group~\cite{ma2026lightweight} distills a
vision--radar teacher into a vision--radar student, so the student still needs
\emph{both} sensors at inference. Park~et~al.~\cite{park2025ckd,
park2025resource} distill a multimodal teacher into a radar-only student, a
\emph{cross-modal} setting, and report a substantial gain over training without
distillation.

That cross-modal evidence, however, comes from \emph{simulated} data (CARLA\slash MATLAB),
\emph{single-step} prediction and a radar input that is a simulated point cloud carrying
per-detection velocity rather than a measured Doppler map. In the vision literature,
distillation across a modality gap is known to be unreliable~\cite{huo2024c2kd}, although
those studies pair \emph{disjoint} modalities, whereas here the teacher also observes the
student's modality. A related line avoids distillation: one multimodal model is trained
with missing-modality masks so that it runs on whichever sensors are
available~\cite{amber2025}. It is untested whether cross-modal transfer survives on real measurements,
at longer horizons and for a radar student that already reads a Doppler-resolved map
(Gap~3); Secs.~\ref{sec:res_pred} and~\ref{sec:res_choices} test it.

\subsection{Positioning of This Work}
\label{sec:rel_position}
None of the surveyed methods predicts several future beams at once from a
Doppler-resolved radar map alone, without CSI. The closest methods use velocity only as a
replacement for the range--angle map in single-slot prediction~\cite{demirhan2022radar},
need the previously confirmed beam~\cite{luo2023radar} or distill within
vision~\cite{ma2025kd}. The other two keep the camera at inference~\cite{ma2026lightweight} or distill
into a radar-only student for single-step prediction on simulated
data~\cite{park2025ckd}. All five report top-$k$ accuracy and none a link-level effect
size in decibels (Sec.~\ref{sec:rel_eval}). This paper measures each prediction's loss
against the best recorded beam, with pass-level intervals, and relates it to accuracy. Because their
evaluations differ from ours (horizon, data or, for~\cite{luo2023radar}, the previously
confirmed beam), we do not place their headline accuracies beside ours. Every
model in Table~\ref{tab:accuracy} is trained and evaluated on one frame-disjoint split.

\looseness=1 Besides the companion paper~\cite{companion_voi}, six cited works share at least one author
with the present paper: \cite{nguyen2026kd},
\cite{ma2025kd}, \cite{ma2026attention}, \cite{ma2026lightweight},
\cite{orimogunje2026beamguard} and \cite{tavakolian2026sub6}. Of these, \cite{ma2025kd} is
also the source of the ADBA metric (DBA averaged over slots), and \cite{ma2025kd}
and~\cite{ma2026attention} are the sources of the backbone, preprocessing and
self-distillation phase adopted here (Sec.~\ref{sec:temporal}).
Two appeared at ICASSP~2026~\cite{ma2026attention, tavakolian2026sub6} and three in IEEE
periodicals~\cite{ma2025kd, nguyen2026kd, orimogunje2026beamguard}, two of them in early access;
the remaining one~\cite{ma2026lightweight} is an arXiv preprint, unrefereed at the time of writing. We disclose authorship and status because these works help define
the present contribution.

\section{Measured-Link Model and Problem Formulation}
\label{sec:system}
\begin{figure}[t]
\centering
\includegraphics{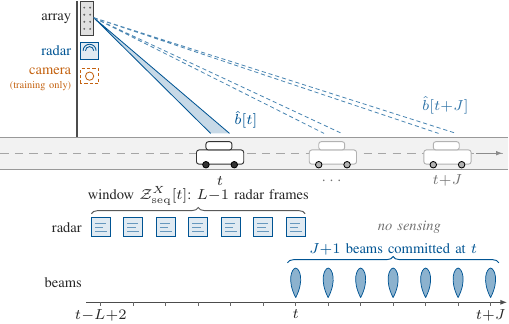}
\caption{System model and window timing. The roadside BS carries a phased array and a
co-located radar; the camera feeds the vision--radar teacher only during offline training
and is absent at deployment. At slot~$t$, the radar-only student reads the window
$\mathcal{Z}^{X}_{\mathrm{seq}}[t]$ of $L\!-\!1$ radar frames and commits the beams
$\hat b[t],\ldots,\hat b[t\!+\!J]$, with no further sensing through slot~$t\!+\!J$; faded vehicles and
dashed beams mark the later slots of that commitment. The measured sweep supplies labels
and link-level evaluation only, never predictor input. Scenario~9 was recorded in daylight and
clear weather throughout.}
\label{fig:system_model}
\end{figure}

We consider a mmWave vehicle-to-infrastructure~(V2I) link in which a
BS equipped with a phased array and co-located sensors serves a
vehicle-mounted user equipment, as illustrated in Fig.~\ref{fig:system_model}.
The dataset measures this link on the uplink: the vehicle's quasi-omnidirectional
transmission is received through the BS's receive codebook. We assume beam reciprocity,
so the best receive beam is also the best transmit beam.
The evaluation uses the dataset's measured sweep over $C\!=\!64$ beams,
not an analytic array pattern. Let $P_c[\tau]$ denote the received power the
dataset records, in linear units, for beam $c\in\{0,\ldots,C-1\}$ at time slot $\tau$;
consecutive slots are consecutive samples, $111$\,ms apart in Scenario~9. The best available beam and the loss incurred by the predicted beam
$\hat b[\tau]$ are
\begin{align}
 b^\star[\tau]&=\argmax_{c}P_c[\tau],
 \label{eq:optimal_beam}\\
 \ell[\tau]&=-10\log_{10}\!\frac{P_{\hat b[\tau]}[\tau]}
 {P_{b^\star[\tau]}[\tau]}.
 \label{eq:measured_loss}
\end{align}
Thus $\ell[\tau]$ is the measured beam-selection loss: path loss, absolute link quality and
fading common to all beams divide out; beam-dependent fading does not.

\subsection{Multimodal Sensing Observations}
\label{sec:observation}
The predictor uses no CSI; the BS instead relies on co-located sensing by an RGB camera and
a frequency-modulated continuous-wave (FMCW) radar, observed over a window of $L$ past
frames. The camera serves the teacher and the vision-only student; the deployed
predictor reads radar alone. The camera yields motion masks $\mathbf{Z}_V[\tau]$
obtained by grayscale conversion, frame differencing and thresholding~\cite{ma2025kd}.
The radar forms its range--Doppler--angle cube by three cascaded fast Fourier transforms
(FFTs), subtracting the
slow-time mean before the Doppler transform. The subtraction suppresses the zero-Doppler component
and with it the response of static scatterers. The radar is the dataset's $77$\,GHz FMCW
unit with four receive channels, $128$ chirps per frame and a maximum velocity of
$56$\,km/h~\cite{demirhan2022radar}, so each of its $128$ Doppler bins spans
$0.24$\,m/s. Supplementary Sec.~S1 lists the other parameters. Averaging the
magnitude of the cube over Doppler, range or angle yields, respectively, the
range--angle (RA), Doppler--angle (DA) and range--Doppler (RD) maps
$\mathbf{R}^{\mathrm{RA}}[\tau]\in\mathbb{R}^{256\times64}$,
$\mathbf{R}^{\mathrm{DA}}[\tau]\in\mathbb{R}^{128\times64}$ and
$\mathbf{R}^{\mathrm{RD}}[\tau]\in\mathbb{R}^{256\times128}$. These sizes are the natural axis lengths before
the fixed-size network inputs are prepared. The reference
configuration uses the two-channel input $[\mathbf{R}^{\mathrm{RA}},\mathbf{R}^{\mathrm{RD}}]$;
DA-only and other combinations are controlled ablations.

Differencing consumes one frame, so $L$ raw frames yield $L\!-\!1$ masks and the radar
maps are aligned to those, giving a sensing window at time~$t$ of $L\!-\!1$ slots,
\begin{equation}
 \begin{aligned}
  \mathcal{Z}^{X}_{\mathrm{seq}}[t]
   &=\{(\mathbf{Z}_V[\tau],\mathbf{R}^{\mathrm{RA}}[\tau],
      \mathbf{R}^{X}[\tau])\}_{\tau=t-L+2}^{t},\\
  &\hspace{28mm}X\in\{\mathrm{DA},\mathrm{RD}\},
 \end{aligned}
  \label{eq:window}
\end{equation}
with the unused inputs omitted for the radar-only student (Fig.~\ref{fig:system_model}) and for the unimodal and single-map ablations. Every encoder-side tensor in
Sec.~\ref{sec:predictor} has length $L\!-\!1$ along its time axis for this reason;
supplementary Fig.~S1 lays the indices out. Three properties of this setup bound what the evaluation can claim. The window
\emph{ends at $t$}, so sensing from slot $t$ is assumed available before that slot's beam
is committed; within-slot acquisition and actuation delays are not modeled. The predictor
receives neither CSI nor any beam index: only the quantities in~\eqref{eq:window} enter
it, so it is CSI-free and feedback-free. The labels are the arg\,max of an exhaustive
$64$-beam sweep rather than a link's own feedback. A predictor holding the last swept
label, called persistence in Sec.~\ref{sec:res_trivial}, is therefore exact at slot $t$
by construction, so comparisons with it are read over the $J$ future slots $t\!+\!1,\dots,t\!+\!J$.

\subsection{Problem Formulation}
\label{sec:problem}
At each window indexed by~$t$, the BS must commit the beams for the current and $J$
future slots from the sensing window alone. Casting this as classification over the
$C\!=\!64$ measured beams, we seek a predictor
\begin{gather}
  f_\theta:\ \mathcal{Z}^{X}_{\mathrm{seq}}[t]\ \longmapsto\
  \{\hat{b}[\tau]\}_{\tau=t}^{t+J},
  \label{eq:predict}\\
  \hat{b}[\tau]=\argmax_{c}\,\pi_c[\tau],
  \label{eq:argmax}
\end{gather}
where $\pi_c[\tau]$ is the predicted probability of beam $c$ at slot $\tau$. The
link-level objective that the evaluation scores is
\begin{equation}
\min_{\theta}\
  \mathbb{E}\!\left[\frac{1}{J+1}\sum_{\tau=t}^{t+J}\ell[\tau]\right],
  \label{eq:problem}
\end{equation}
where the expectation is over sensing windows and vehicle passes. Like the surveyed
predictors, the models are trained as classifiers on the arg\,max label, here with a focal loss
(Sec.~\ref{sec:loss}). The recorded powers could serve as soft targets, but we keep the
one-hot objective for every model and leave power-weighted training to future
work. Sec.~\ref{sec:res_resolution} therefore tests how well top-$1$
accuracy tracks the objective in~\eqref{eq:problem}. The protocol is windowed and
open-loop: each window is evaluated on recorded sensing and labels, and a wrong beam has
no effect on later windows. The results are therefore an optimistic benchmark, not an
estimate of closed-loop performance.

\section{Doppler-Augmented Predictor and Distillation Testbed}
\label{sec:predictor}
\begin{figure*}[t]
\centering
\includegraphics{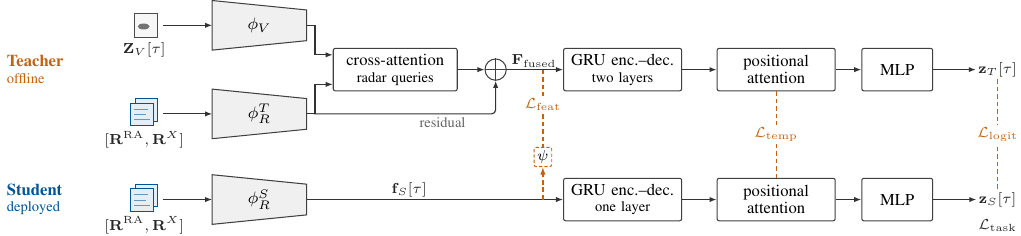}
\caption{Controlled teacher--student testbed. The radar-only student, trained in Phase~3, is the
deployed predictor; the reference student trains on $\mathcal{L}_{\mathrm{task}}$ alone, with all three
KD weights zero. The vision--radar teacher is trained offline (Phases~1--2). Solid arrows are
forward paths; dashed orange links are distillation signals, used only in the distillation
experiments. Feature alignment matches $\mathbf{F}_{\mathrm{fused}}$ to the student
features projected by $\psi$, temporal-attention alignment matches teacher and student attention
maps, and logit distillation matches $\mathbf{z}_T$ and $\mathbf{z}_S$. The two radar
encoders ($\phi_R^{T}$, $\phi_R^{S}$) share the Doppler-augmented architecture but are optimized separately; the
student has a one-layer gated recurrent unit (GRU) and no vision or fusion branch. Here
$X\!\in\!\{\mathrm{DA},\mathrm{RD}\}$.}
\label{fig:framework}
\end{figure*}

\looseness=1 This section defines the Doppler-augmented radar-only predictor and the high-capacity
vision--radar teacher used to test whether cross-modal distillation improves the predictor
(Fig.~\ref{fig:framework}). The reference student is trained without distillation;
the teacher and transfer losses are experimental apparatus. The radar encoder ingests
motion information through a Doppler-resolved map (Sec.~\ref{sec:doppler}), the input
choice with the dominant measured effect in this long-term radar-only predictor family.

In words, the radar-only predictor works as follows (Fig.~\ref{fig:framework}). For each of the
seven observed frames, a compact convolutional network turns the range--angle map and the Doppler-resolved map
into one feature vector. That vector encodes where the moving vehicle is and how fast it approaches
or recedes. A recurrent network reads these vectors in order, so it can follow how position
and velocity evolve. An attention-based decoder then turns the recurrent state into a beam distribution
for each slot of the horizon, the current one and six ahead. The teacher adds a camera
branch and fuses it with the radar features. During training only, its outputs, features and attention maps
serve as targets for the student, which at inference reads radar alone.

\subsection{Doppler-Augmented Modality Encoders}
\label{sec:doppler}
For each of the $L\!-\!1$ past frames, modality-specific convolutional neural networks
(CNNs) produce
$D$-dimensional embeddings. The vision branch encodes the motion mask,
$\mathbf{f}_V[\tau]=\phi_V(\mathbf{Z}_V[\tau];\theta_V)$, where $\phi_V$ is a five-block CNN with a fully connected
head ($1.71$\,M parameters). Whereas prior
radar-only pipelines read a single radar representation, the radar branch here
ingests the range--angle map and a second, Doppler-resolved map
as a \emph{two-channel} input,
\begin{equation}
  \mathbf{f}_R[\tau]=\phi_R\big([\mathbf{R}^{\mathrm{RA}}[\tau],\mathbf{R}^{X}[\tau]];\theta_R\big)\in\mathbb{R}^{D},
  \quad X\in\{\mathrm{DA},\mathrm{RD}\},
  \label{eq:radar_enc}
\end{equation}
so a single encoder jointly extracts \emph{position} and \emph{radial-velocity}
features. When $X\!=\!\mathrm{DA}$, stacking lets one encoder read both maps along angle, their
only shared axis, which defines the beam index. The results do not support this
shared-axis hypothesis: the reference pairing is range--Doppler, which instead shares the range axis
with the range--angle map, yet performs the same as the Doppler--angle pairing to within
$0.02$\,dB. Separate input branches per map do no better (Sec.~\ref{sec:setup}).

Two configurations are therefore relevant. A single Doppler--angle map is the
leaner deployment candidate, one input map rather than two. Its deficit against the reference pairing
is not resolved ($0.03$\,dB in the pairing's favor on $[-0.09, +0.15]$), so
equivalence is not established either. We report the range--Doppler pairing throughout as the
\emph{reference configuration} so that all comparisons use the same input reference. It
became the reference after a three-seed pilot, read on the test split, had placed it
$1.4$ percentage points (pp) above the Doppler--angle pairing. At ten seeds the two are indistinguishable
(Sec.~\ref{sec:res_pred}). The Doppler--angle pairing gains as much over range--angle
alone ($0.26$\,dB on $[0.13, 0.39]$, against $0.28$\,dB), so the choice did not create the
Doppler gain. In top-$1$ accuracy, the distillation null of Sec.~\ref{sec:res_ablation} also
holds for the Doppler--angle pairing (Table~\ref{tab:accuracy}).

\looseness=1 Teacher and student use the same radar-encoder architecture with separate, not
weight-tied, parameters. This keeps their comparison interpretable, since a teacher
advantage cannot be attributed to a larger radar encoder; the cost is a cap on what the
teacher can extract from radar. Per-frame embeddings are stacked into
$\mathbf{F}_V,\mathbf{F}_R\!\in\!\mathbb{R}^{(L-1)\times D}$. The radar branch is
deliberately compact, three convolutional blocks followed by global pooling, so that the
radar-only student can use the same architecture. The two-channel input adds only the
first convolution's input weights over a range--angle-only encoder. Supplementary
Table~S1 details both encoders.

\subsection{Cross-Attention Fusion in the Teacher}
\label{sec:fusion}
The teacher fuses modalities by multi-head cross-attention in which \emph{radar
poses the queries and vision the keys/values}, with an output projection and a residual
on the radar branch,
\begin{equation}
  \mathbf{F}_{\mathrm{fused}}=\mathrm{LayerNorm}\big(
    \mathrm{MHA}(\mathbf{F}_R,\mathbf{F}_V,\mathbf{F}_V)+\mathbf{F}_R\big),
  \label{eq:fused}
\end{equation}
where $\mathrm{MHA}$ is standard multi-head attention over $N_h\!=\!4$ heads of width
$D/N_h$ with $D\!=\!64$. The decoder-to-encoder attention below uses eight heads (all
other dimensions are listed in supplementary Sec.~S2). The residual is the design choice: up to layer normalization, \eqref{eq:fused} adds
an attention update to $\mathbf{F}_R$, so the fused representation stays
anchored to the radar features. A concatenation $[\mathbf{F}_V,\mathbf{F}_R]$, by
contrast, places the teacher in a space whose vision half a radar-only student cannot
reproduce~\cite{park2025ckd}, an argument that holds at the feature level only. The concatenation
teacher of Sec.~\ref{sec:res_fusion} tests this hypothesis and does not support it.

\subsection{Temporal Model With Positional Attention}
\label{sec:temporal}
The fused sequence is processed by a gated recurrent unit (GRU) encoder--decoder. The encoder summarizes the
$L\!-\!1$ past frames; the decoder is then run for $J\!+\!1$ further steps on zero-padded
inputs, carrying the encoder's final hidden state forward. Since the decoder inputs are
all zero, no prediction is fed back between steps and the recurrence acts as a set of
learned per-horizon queries rather than an autoregressive decoder. Because the decoder steps differ only through the recurrence,
learned positional encodings $\mathbf{P}_{\mathrm{enc}}$ and $\mathbf{P}_{\mathrm{dec}}$ are
added to the encoder and decoder states $\mathbf{H}_{\mathrm{enc}}$ and $\mathbf{H}_{\mathrm{dec}}$
before a decoder-to-encoder attention,
\begin{equation}
  \mathbf{A}^{(i)}=\mathrm{softmax}\!\Big(
  \frac{\mathbf{Q}^{(i)}\mathbf{K}^{(i)\top}}{\sqrt{d_k'}}\Big)
  \in\mathbb{R}^{(J+1)\times(L-1)},
  \label{eq:attn}
\end{equation}
where head $i$ of eight projects queries from
$\mathbf{H}_{\mathrm{dec}}\!+\!\mathbf{P}_{\mathrm{dec}}$ and keys from
$\mathbf{H}_{\mathrm{enc}}\!+\!\mathbf{P}_{\mathrm{enc}}$ to $d_k'\!=\!D/8\!=\!8$ dimensions. Row~$j$ is the
importance assigned to the past frames when predicting slot $t\!+\!j$, so the map can express
which past frame informs which future slot. The refined decoder states pass through a
multilayer perceptron (MLP) head to yield logits. The student shares this structure, so the teacher's attention map is
available as an optional distillation target. The convolutional--recurrent backbone with
attention over the recurrent states, the motion-mask preprocessing and the
self-distillation phase follow the vision-only pipelines of~\cite{ma2025kd,
ma2026attention}, where multi-head attention acts on the recurrent outputs. The
decoder-to-encoder form of that attention with learned positional encodings, the
Doppler-resolved radar input, the radar-only student and its cross-modal distillation
are added here.

\subsection{Radar-Only Student}
\label{sec:student}
The student reads radar alone in training and inference. It uses the same
two-channel radar-encoder architecture~\eqref{eq:radar_enc}, a one-layer GRU and the
same positional attention~\eqref{eq:attn}. It produces features
$\mathbf{f}_S[\tau]$, attention maps $\mathbf{A}_S^{(i)}$ dimensionally identical to the
teacher's, and logits $\mathbf{z}_S[\tau]$ whose softmax gives the probabilities
$\pi_c[\tau]$ of~\eqref{eq:argmax}. It carries $92.5$\,k parameters at inference ($96.6$\,k in training, the
difference being the projector $\psi$ of Sec.~\ref{sec:loss}) against the teacher's
$1.84$\,M, a factor of about $20$. The reduction comes from omitting the vision branch
and the fusion and from a shallower recurrence; supplementary Table~S2 gives the
per-component counts. Computation and latency do not shrink in proportion: per call the
student needs more multiply--accumulates than the vision-only student
(Sec.~\ref{sec:res_pred}). Its practical advantage is that it needs \emph{no
camera or vision pipeline at inference}; whether that yields robustness to darkness and
weather is not established here (Sec.~\ref{sec:res_pred}).

\subsection{Composite Distillation Loss}
\label{sec:loss}
The ground-truth beam is $b^\star[\tau]$ of~\eqref{eq:optimal_beam}. The loss combines a task term with
three candidate distillation terms at the output, feature and temporal levels; the
ablation in Sec.~\ref{sec:res_ablation} measures whether any of them helps the
Doppler-augmented student.

\emph{Task loss.} A focal loss~\cite{lin2017focal} addresses beam-class imbalance,
$\mathcal{L}_{\mathrm{task}}=\frac{1}{J+1}\sum_{\tau}-\alpha_f(1-\pi_{b^\star}[\tau])^{\gamma_f}\log \pi_{b^\star}[\tau]$,
with $(\alpha_f,\gamma_f)\!=\!(1,2)$, averaged, like every term below, over the windows
of a batch.

\emph{Logit distillation} applies a temperature-$\Gamma$ Kullback--Leibler (KL) divergence
between softened outputs,
\[
  \mathcal{L}_{\mathrm{logit}}=\frac{\Gamma^2}{J+1}\sum_{\tau}D_{\mathrm{KL}}
  \big(\sigma(\mathbf{z}_T[\tau]/\Gamma)\,\|\,\sigma(\mathbf{z}_S[\tau]/\Gamma)\big),
\]
with $\sigma$ the softmax, $\mathbf{z}_T[\tau]$ the teacher's logits and the factor $\Gamma^2$ restoring the gradient scale. \emph{Feature alignment} maps student features
to the teacher's fused representation through a linear projector $\psi$, discarded at inference,
and penalizes the mean squared error over frames and features,
$\mathcal{L}_{\mathrm{feat}}=\frac{1}{(L-1)D}\|\mathbf{F}_{\mathrm{fused}}-\psi(\mathbf{F}_S)\|_F^2$ with $\mathbf{F}_S$ the student's stacked features. It asks the student's
radar features to predict the teacher's vision-informed ones. \emph{Temporal-attention alignment} matches the teacher's attention
maps $\mathbf{A}_T^{(i)}$ row by row, $\mathcal{L}_{\mathrm{temp}}=\frac{1}{8(J+1)}\sum_{i,j}D_{\mathrm{KL}}([\mathbf{A}_T^{(i)}]_{j,:}\,\|\,[\mathbf{A}_S^{(i)}]_{j,:})$, which
asks the student to weight the past frames as the teacher does. We match head $i$
to head $i$, an arbitrary pairing we did not vary. The total objective is
\begin{equation}
  \mathcal{L}=\mathcal{L}_{\mathrm{task}}+\lambda_1\mathcal{L}_{\mathrm{logit}}
  +\lambda_2\mathcal{L}_{\mathrm{feat}}+\lambda_3\mathcal{L}_{\mathrm{temp}}.
  \label{eq:total}
\end{equation}
The reference student sets all three weights to zero by definition
($\lambda_1\!=\!\lambda_2\!=\!\lambda_3\!=\!0$) and trains on $\mathcal{L}_{\mathrm{task}}$
alone. The ablation of Sec.~\ref{sec:res_ablation} adds the three terms cumulatively.

\subsection{Training Procedure}
\label{sec:training}
Training proceeds in three phases. \emph{Phase~1}: the
teacher is trained from scratch with $\mathcal{L}_{\mathrm{task}}$.
\emph{Phase~2}: it is refined by self-distillation~\cite{ma2025kd,mobahi2020self}
using its own softened Phase-1 logits.
\emph{Phase~3}: with the teacher frozen, the student is trained with~\eqref{eq:total};
both networks see identical radar sequences while only the teacher additionally
receives vision. At inference, the radar-only student is deployed alone.

\section{Experimental Setup and Prediction Accuracy}
\label{sec:results}
This section specifies the protocol and reports the empirical ordering of the predictors,
the principal contrasts and a comparison with nonlearned references.
Sec.~\ref{sec:res_resolution} then justifies received power as their common scale, and
Sec.~\ref{sec:res_choices} asks what explains the two contrasts.

\subsection{Experimental Setup}
\label{sec:setup}
\textbf{Dataset and splits.} We use Scenario~9 of the real-world DeepSense~6G
dataset~\cite{10144504}: a V2I setting with time-synchronized
RGB images, FMCW radar and $60$\,GHz beam-training measurements over a
$C\!=\!64$ narrow-beam codebook. Its $5{,}964$ samples form $136$ discrete
vehicle passes, recorded at $9.0$\,Hz ($111$\,ms between frames). An $L\!=\!8$ sliding window
yields $4{,}060$ sequences, each predicting the current and $J\!=\!6$ future slots, so the
$L\!-\!1\!=\!7$ observation frames and the horizon each span $0.67$\,s. We split
chronologically \emph{at pass boundaries}, so no frame is shared across splits. The
first $91$ passes ($2{,}787$ sequences) train, the next $17$ ($499$) validate and the last
$28$ ($774$) test; early stopping and checkpoint selection use the validation passes alone.
We call one model configuration trained over its seeds, ten unless otherwise stated, an
\emph{arm}. Scenario~31, a second BS site $110$\,m along the same street, is used only to
replicate the Doppler contrast (supplementary Sec.~S6 gives its splits and protocol).

\textbf{Preprocessing.} Vision is reduced to $224\!\times\!224$ binary motion masks. Each radar
map of Sec.~\ref{sec:observation} is log-compressed and prepared as a fixed
$128\!\times\!64$ network input with no further normalization, so no statistic is
estimated from any split (supplementary Sec.~S1 gives the crops). The common input shape
does not mean that the axes match: rows are range in $\mathbf{R}^{\mathrm{RA}}$ and
$\mathbf{R}^{\mathrm{RD}}$ but Doppler in $\mathbf{R}^{\mathrm{DA}}$, which matters for the
controls of Sec.~\ref{sec:res_doppler}. All learned arms use the same stored maps and the
same preprocessing.

\textbf{Baselines and ablations.} We compare the radar-only student with and without
distillation, a \emph{vision-only} student and the vision--radar teacher as a
high-capacity reference, all trained and evaluated on the same split with the same budget
and checkpoint rule. The vision-only student has the radar-only student's temporal architecture, with
the vision encoder $\phi_V$ in place of the radar encoder, trained on
$\mathcal{L}_{\mathrm{task}}$ alone. A Doppler--angle or range--Doppler teacher or student
pairs $\mathbf{R}^{\mathrm{RA}}$ with the map it is named after; a range--angle one uses
$\mathbf{R}^{\mathrm{RA}}$ alone. The teacher row of Table~\ref{tab:accuracy} is the
Phase\nobreakdash-2, self-distilled RA$+$DA teacher; the RA$+$RD students are distilled from an
RA$+$RD teacher trained the same way (supplementary Sec.~S10). We ablate the distillation loss, the Doppler-resolved
input and the teacher fusion topology (Secs.~\ref{sec:res_pred} and~\ref{sec:res_choices}). None of these
baselines reproduces a published system: the range--angle-only row is the testbed's
encoder with the Doppler-resolved map removed, not an implementation
of~\cite{demirhan2022radar} or~\cite{luo2023radar}.

\textbf{Fairness checks.} A two-channel encoder carries $144$ more parameters, all in the
first convolution; Sec.~\ref{sec:res_pred} shows that this capacity does not explain the
Doppler gain. Separate input branches per map do no better than the shared encoder, and
the range--angle arm is worse without the mean subtraction, so the Doppler gain is measured
against the better range--angle input (supplementary Sec.~S5).

\looseness=-1 \textbf{Metrics.} Each window predicts $J\!+\!1\!=\!7$ slots, so the test set has $774\times7$
evaluated slots, from overlapping windows. Prediction quality is measured by the average
top-$k$ accuracy (ATop\nobreakdash-$k$, $k\!\in\!\{1,3,5\}$) and ADBA
($\Delta\!=\!5$)~\cite{ma2025kd}, which scores each slot
$1-\min_{k\le3}\min(|\hat b_k-b^\star|/\Delta,1)$ over its top three predictions. Both are
averaged over the $7$ slots for comparability with~\cite{ma2025kd}. Link quality is the
per-slot beam-selection loss~\eqref{eq:measured_loss}, and the comparisons rest on it
(Sec.~\ref{sec:res_resolution}). ADBA's credit spans $14\%$ of this oversampled codebook
and compresses the differences between arms, so we report it for comparability only. Where
a comparator is exact at slot $t$ by construction, we read $t\!+\!1$ to $t\!+\!6$ instead and
say so; reading Tables~\ref{tab:accuracy} and~\ref{tab:capture} over those six slots reorders only near-tied arms (supplementary
Sec.~S1).

\textbf{Statistics.} Because sliding windows overlap within a vehicle pass, the effective
sample size is the $28$ test passes rather than the $774$ windows. Every interval is a
$95\%$ cluster-bootstrap interval that resamples passes (as whole clusters of windows,
shared by the arms of a contrast) and seeds within each arm. Supplementary
Sec.~S3 states the procedure in full and checks the two central contrasts with other
constructions. We
treat sub-point accuracy differences as noise unless the interval excludes zero. With
some forty intervals read at the $95\%$ level without multiplicity correction, those that
clear zero by a small margin are indicative only.

\textbf{Pre-specification.} No Scenario~9 contrast is strictly confirmatory. Every arm
reads test passes that earlier arms used, the reference was chosen on a pilot read on them
(Sec.~\ref{sec:doppler}), and both headline effects were first seen there in single-seed
runs. Scenario~9 intervals therefore capture sampling, not selection, uncertainty. The arms
fall into three tiers. \emph{Specified} arms were fixed before their own results were read.
\emph{Pre-specified} arms follow a written decision rule in a dated file (released with the code): the Scenario~31 replication, the $\Gamma\!=\!1$ rerun, the
distillation from two further teachers and the Doppler-centroid test of
Sec.~\ref{sec:res_doppler}. \emph{Exploratory} arms are claimed only as bounds. Here,
pre-specified means that the rule preceded the analysis, not that the data were unseen.
Only the Scenario~31 contrast had not been read on its test passes (supplementary Table~S3
dates every arm and gives its tier).

\textbf{Implementation.} Models are implemented in PyTorch; supplementary Sec.~S2 gives the
optimizer, schedule, loss constants and checkpoint rule (validation ADBA). Distillation
uses temperature $\Gamma\!=\!3$ and unit weights, a convention; a pre-specified ten-seed
rerun at the validation-best $\Gamma\!=\!1$ changes nothing (supplementary Sec.~S10). Every
ten-seed arm uses the same seeds; tables report the mean $\pm$ sample standard deviation
(std) over seeds. The split indices, the test-set predictions and the preprocessing, training,
evaluation and bootstrap scripts are available at \url{https://github.com/jeongwankang95/radar-beam-prediction}.

\subsection{Long-Term Prediction Accuracy}
\label{sec:res_pred}
\begin{figure}[t]\centering
\includegraphics{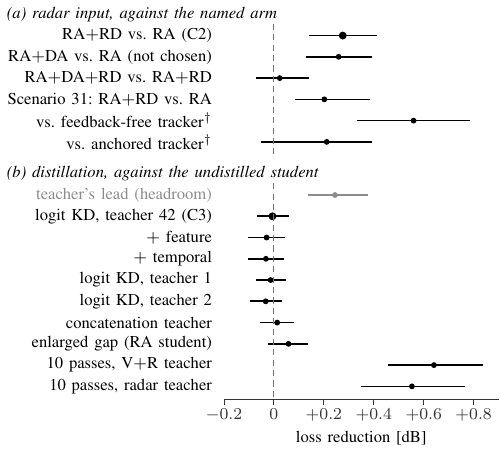}
\caption{Link-level contrasts as loss reduction in decibels, positive when the first-named arm loses
less, with $95\%$ intervals that resample test passes and seeds. (a) The two Doppler pairings
against range--angle alone, all three maps against the reference configuration, the Scenario~31
replication on that site's test passes (corrected labels; the pre-specified readout is
$+0.14$\,dB on $[+0.05, +0.26]$) and the reference student against the
road-coordinate tracker, feedback-free and anchored ($^{\dagger}t\!+\!1$ to $t\!+\!6$). (b) Distillation against the
undistilled student (RA$+$RD teachers named by seed), including the ten-pass student of the learning
curve, with the source teacher's lead (gray) for scale.}
\label{fig:forest}
\end{figure}

One input effect dominates the radar-only students of Table~\ref{tab:accuracy}: that of a
Doppler-resolved map. Paired with the range--angle map, it lifts the undistilled student by
$4.0$ to $4.6$\,pp of ATop-1 and, for the reference range--Doppler pairing, lowers the mean
beam-selection loss~\eqref{eq:measured_loss} by $0.28$\,dB on $[0.14, 0.41]$ (Fig.~\ref{fig:forest}). The gain comes
from the Doppler content, not from the extra input: retrained with an all-zero second map,
so that the graph and every parameter match, the range--angle student changes its loss by
less than $0.01$\,dB.

\begin{table}[t]\centering
\caption{Long-term accuracy (\%), averaged over $J\!+\!1\!=\!7$ slots, as mean$\pm$std over
ten training seeds ($^{\star}$reference, undistilled). The distillation rows add their loss
terms cumulatively to the student they name; Sec.~\ref{sec:setup}
identifies the teacher row, and supplementary Sec.~S10 names the distillation sources.}
\label{tab:accuracy}\footnotesize
\setlength{\tabcolsep}{2.5pt}
\begin{tabular}{@{}lccccc@{}}
\toprule
\textbf{Model / configuration} & \textbf{Mod.} & \textbf{ATop-1} & \textbf{ATop-3} & \textbf{ATop-5} & \textbf{ADBA}\\
\midrule
Vision--radar teacher        & V$+$R & $45.8{\scriptstyle\pm1.2}$ & $81.9{\scriptstyle\pm1.3}$ & $93.5{\scriptstyle\pm0.7}$ & $94.7{\scriptstyle\pm0.6}$\\
Vision-only                  & Vis.  & $44.1{\scriptstyle\pm1.8}$ & $79.5{\scriptstyle\pm2.6}$ & $92.7{\scriptstyle\pm2.0}$ & $94.4{\scriptstyle\pm0.7}$\\
\midrule
Range--angle only (RA)       & Rad. & $37.3{\scriptstyle\pm1.2}$ & $69.6{\scriptstyle\pm2.4}$ & $86.7{\scriptstyle\pm2.3}$ & $89.9{\scriptstyle\pm0.8}$\\
\midrule
\multicolumn{6}{@{}l@{}}{\emph{Doppler \textbf{replacing} the range--angle map}}\\
\quad DA only                & Rad. & $41.8{\scriptstyle\pm1.6}$ & $74.3{\scriptstyle\pm3.5}$ & $88.9{\scriptstyle\pm2.3}$ & $92.3{\scriptstyle\pm0.7}$\\
\quad RD only                & Rad. & $39.8{\scriptstyle\pm1.1}$ & $72.2{\scriptstyle\pm2.6}$ & $88.0{\scriptstyle\pm1.9}$ & $91.6{\scriptstyle\pm0.5}$\\
\midrule
\multicolumn{6}{@{}l@{}}{\emph{Doppler \textbf{augmenting} it}}\\
\quad RA$+$DA                & Rad. & $41.3{\scriptstyle\pm0.9}$ & $76.7{\scriptstyle\pm0.8}$ & $90.4{\scriptstyle\pm0.5}$ & $92.6{\scriptstyle\pm0.5}$\\
\quad \textbf{RA$+$RD}$^{\star}$ & Rad. & $41.9{\scriptstyle\pm1.0}$ & $76.6{\scriptstyle\pm2.2}$ & $90.7{\scriptstyle\pm1.4}$ & $92.7{\scriptstyle\pm0.6}$\\
\quad RA$+$DA$+$RD & Rad. & $42.3{\scriptstyle\pm1.1}$ & $77.1{\scriptstyle\pm2.2}$ & $90.6{\scriptstyle\pm1.4}$ & $92.9{\scriptstyle\pm0.6}$\\
\quad RA$+$RD $+$ logit & Rad. & $41.4{\scriptstyle\pm1.1}$ & $77.0{\scriptstyle\pm1.1}$ & $91.0{\scriptstyle\pm0.5}$ & $92.5{\scriptstyle\pm0.4}$\\
\quad \qquad $+$ feature & Rad. & $41.2{\scriptstyle\pm0.9}$ & $76.3{\scriptstyle\pm1.6}$ & $90.7{\scriptstyle\pm1.0}$ & $92.3{\scriptstyle\pm0.4}$\\
\quad \qquad $+$ temporal & Rad. & $41.0{\scriptstyle\pm0.9}$ & $76.2{\scriptstyle\pm1.6}$ & $90.6{\scriptstyle\pm1.0}$ & $92.3{\scriptstyle\pm0.5}$\\
\midrule
\multicolumn{6}{@{}l@{}}{\emph{Other distillation variants (RA$+$DA student)}}\\
\quad logit & Rad. & $41.4{\scriptstyle\pm0.9}$ & $76.8{\scriptstyle\pm1.1}$ & $90.7{\scriptstyle\pm0.8}$ & $92.6{\scriptstyle\pm0.6}$\\
\quad \qquad $+$ feature & Rad. & $40.5{\scriptstyle\pm1.3}$ & $75.3{\scriptstyle\pm1.5}$ & $90.2{\scriptstyle\pm1.0}$ & $92.2{\scriptstyle\pm0.6}$\\
\quad \qquad $+$ temporal & Rad. & $40.7{\scriptstyle\pm1.3}$ & $75.5{\scriptstyle\pm1.4}$ & $90.4{\scriptstyle\pm1.0}$ & $92.2{\scriptstyle\pm0.5}$\\
\bottomrule
\end{tabular}
\end{table}

What matters is the presence of a Doppler-resolved map, not how it enters. Every
radar-only configuration containing one lands within $0.13$\,dB of every other, while the
range--angle-only student, at $0.91$\,dB, is the sole outlier (Table~\ref{tab:capture}).
Neither the choice of projection, nor keeping the range--angle map beside the better single
map, nor stacking both projections changes the loss measurably (supplementary Sec.~S14).
Sec.~\ref{sec:res_doppler} asks what in that map the encoder uses.

The Doppler gain replicates at a second BS site, $110$\,m along the same street.
Under a decision rule written before the contrast was read, the range--Doppler pairing
gains $+0.14$\,dB there, and $+0.20$\,dB after a label correction and retraining
(Fig.~\ref{fig:forest}; supplementary Sec.~S6). What replicates is the Doppler contrast, not
the predictor's lead over persistence. At that site, persistence (Sec.~\ref{sec:res_trivial}) beats both
radar arms over $t\!+\!1$ to $t\!+\!6$, because the optimal beam stands still on $40\%$ of frames against $15\%$
here.

Cross-modal distillation, by contrast, adds no measurable gain on the reference
configuration. With the full training set, the distilled and undistilled RA$+$RD
students differ by $-0.01$\,dB of received power ($[-0.07, +0.06]$). The source teacher had room to
transfer: it leads the student by $0.24$\,dB, a headroom comparable to the Doppler gain, yet
the share that distillation recovers has an interval excluding more than a quarter of it.
That headroom is mostly visual. The vision-only student alone accounts for $0.19$\,dB of that
lead, and inside the teacher, radar adds nothing measurable to vision. The null qualifies
rather than contradicts the benefit reported in~\cite{park2025ckd}, which uses simulated
data, single-step prediction and a point-cloud radar student. Sec.~\ref{sec:res_fusion} tests
alternative explanations, from the fusion topology to the teacher--student gap.

On this clear-weather, daytime scenario, the camera is the stronger sensor at full size:
the motion-mask vision-only student ($1.76$\,M parameters) leads every radar arm and comes close
to the teacher (Table~\ref{tab:accuracy}). At the radar-only student's size the ordering reverses.
A vision-only student cut to that size trails the radar-only student by $0.34$\,dB (exploratory, one
matched design; supplementary Sec.~S5). The radar-only pipeline therefore saves the camera,
its processing path and most of the model, with about twenty times fewer parameters than the
teacher or the vision-only student. It does not save arithmetic: it needs $156$ million
multiply--accumulates per call, against $137$ million for the vision-only student.
Distillation is not the source of the parameter reduction. Latency is proportional to neither count: on
a laptop-class CPU, a call takes $2.4$\,ms for the radar-only student and $4.4$\,ms for the
vision-only student~\cite{companion_voi}. The one same-road day/night pair
DeepSense~6G contains, Scenarios~32 and~33, was inconclusive: the radar-only student collapses
there in daylight to the most frequent beam, and the pair cannot separate lighting from
site shift (supplementary Sec.~S7).

\subsection{Comparison With Nonlearned References}
\label{sec:res_trivial}
The reference student beats five nonlearned predictors by a wide margin. Two are
oracle-fed offline references that consume the optimal-beam labels of the window's eight raw
frames, which a deployed system would have to acquire by sweeping. \emph{Persistence} holds
the last measured beam, and \emph{linear} extrapolates a least-squares fit to the eight.
\emph{Constant} uses the most frequent beam of the training split. Two radar readers map the
peak of the Doppler--angle map's per-angle profile to a beam index. One extrapolates the
peak's trajectory from the radar alone, and the other, \emph{anchored}, applies its rate to
the last measured beam. Supplementary Table~S7 tabulates every reference. Over the horizon
proper ($t\!+\!1$ to $t\!+\!6$), the most accurate of the five, the anchored reader, reaches
$18.6\%$ and loses $1.68$\,dB, against $41.8\%$ and $0.65$\,dB for the reference student.

Both radar readers discard the radial velocity that motivates a Doppler-resolved input,
so we also built a tracker that keeps it. It is an extended Kalman filter in road coordinates that
estimates the along-road position and speed of the strongest moving return from its range,
angle centroid and signed Doppler centroid. It is the kinematic core of radar-assisted
predictive beamforming~\cite{liu2020radar, yuan2021bayesian} without a detection front end,
fitted on the training passes (supplementary Sec.~S8).

The student leads the tracker's better feedback-free variant (without the radial velocity) by
$20.8$\,pp and $0.56$\,dB. It leads the anchored variant by $17.9$\,pp, but at the link by only
$0.21$\,dB on $[-0.05, +0.39]$, an interval including zero. The radial velocity
does not help the filter, because the along-road speed barely varies here. Anchoring consumes
the last measured beam, a feedback quantity a deployed link would hold; the predictor under
study is feedback-free by construction, and the causal value of feedback is the companion
paper's question~\cite{companion_voi}.

\begin{figure}[t]\centering
\includegraphics{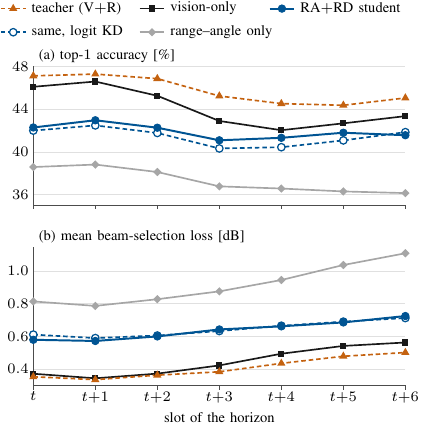}
\caption{Top-$1$ accuracy (a) and mean beam-selection loss (b) at each slot of the horizon
(Scenario~9 test passes, means over ten seeds). Unlike persistence, exact at $t$ but at
$12.5\%$ one slot later and $0.0\%$ at $t\!+\!6$ (off scale), the learned predictors keep most
of their accuracy across the horizon while their loss grows. The Doppler-resolved student
stays ahead of range--angle alone at every slot, and distillation tracks the undistilled
student.}
\label{fig:perslot}
\end{figure}

\looseness=1 The predictor's value arises once the last measured beam has gone stale. Persistence is
exact at the current slot but reaches only $12.5\%$ one slot later and $0.0\%$ at
$t\!+\!6$, while the student stays between $41.1$ and $43.0\%$ at every slot (Fig.~\ref{fig:perslot}). A learned
encoder thus outperforms every nonlearned baseline constructed here in accuracy, and every
feedback-free one at the link. The window itself is worth about two points of accuracy: fed
only its last frame, the student loses $2.2$\,pp (exploratory; supplementary Sec.~S5).

\section{Received Power as the Common Scale for Comparing Predictors}
\label{sec:res_resolution}
Top-$1$ accuracy is a coarse scale on this codebook because the beam peak is flat. Across
all $5{,}964$ measured $64$-beam power vectors, the second-best beam is within $0.21$\,dB of
the best at the median and is an immediate neighbor $87.4\%$ of the time (supplementary
Sec.~S14 shows that the labels are nonetheless stable). A flat peak lets the metric fail in
either direction: scoring near and far misses alike compresses large power differences, and a
predictor systematically one index off loses accuracy at little power cost, inflating small
ones. The preceding section, like most of the works of Sec.~\ref{sec:related}, leads with
accuracy. Whether the one- and two-point differences on which this literature settles
design choices mean anything in received power has to be measured. We measure it with the
power vectors that accompany the labels.

\begin{figure}[t]\centering
\includegraphics{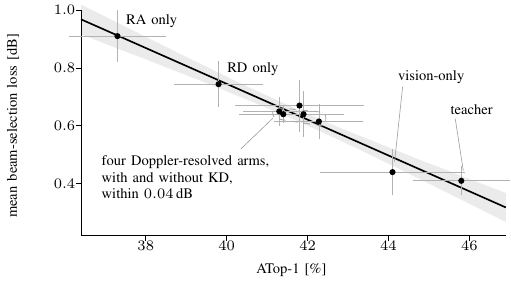}
\caption{Mean beam-selection loss against top-$1$ accuracy for the nine predictors of
Table~\ref{tab:capture}, each point a mean over ten training seeds with $\pm$std bars on
both axes. The line is the least-squares fit ($-0.062$\,dB per point, $R^{2}\!=\!0.97$) with
its pointwise $95\%$ confidence band for this test set ($\pm0.02$\,dB near the centroid,
$\pm0.05$ at the plot edges). Resampling passes and seeds gives the slope interval $[-0.077, -0.047]$. One slope fits the whole range, not only its extremes. The boxed
cluster shows the distillation null at this scale: the four two- and three-map
Doppler-resolved configurations occupy a region $0.04$\,dB deep against the
$0.51$\,dB spread of the nine predictors.}
\label{fig:calib}
\end{figure}

\looseness=1 Accuracy orders the predictors almost as received power does, but a point of it is worth little received power. For each of the nine
predictors of Table~\ref{tab:capture} we compute the \emph{power-capture ratio} $\rho[\tau]$,
the received power its top-$1$ beam collects divided by the optimal beam's, over all seven
slots of the $774$ test sequences. Every Scenario~9 mean-loss figure is the
decibel-domain mean $\mathbb{E}[-10\log_{10}\rho]$, computed from the measured powers without
an array model. Of the $36$ predictor pairs, $34$ are ordered identically by accuracy and by
received power, and the two inversions are within half a point, so the metric is informative
down to about a point. The $8.5$\,pp spread
of the nine predictors corresponds to only $0.51$\,dB of mean loss, and the fitted descriptive slope is
$-0.062$\,dB per percentage point (Fig.~\ref{fig:calib}). The slope fits the observed range,
not only its ends, and no single arm determines it.

The slope is descriptive, not a calibration constant. A more accurate predictor misses less
often and, when it misses, tends to land closer. Because a hit costs nothing,
$\mathbb{E}[\ell]=(1-A)\,\mathbb{E}[\ell\mid\mathrm{miss}]$, with $A$ the top-$1$ accuracy.
The cost per miss falls from $1.46$\,dB for the range--angle-only student to $0.75$ for the
teacher. Refitted on other subsets of arms and seeds, including Scenario~31's, the slope
stays between $-0.044$ and $-0.059$ (supplementary Sec.~S9). Its sign and order of magnitude
recur, but its value moves with the predictor set. A predictor can also depart from it: the
distilled ten-pass student of Sec.~\ref{sec:res_fusion} gains $0.64$\,dB for one point,
because distillation moves its misses closer rather than making them rarer. For the full-data variants
compared in this paper, the one- or two-point ATop-1 differences are worth about a tenth of
a decibel or less.

The decibel differences come mostly from the tail of the loss (Table~\ref{tab:capture}). With the
Doppler-resolved map, the share of slots losing more than $3$\,dB halves, from $8.2\%$ to
$4.2\%$, while distillation moves no tail statistic measurably. Removing the camera (vision-only against the reference student) costs only
$0.19$\,dB in the mean but $3.2$\,dB at the $99$th percentile. The mean therefore understates
both what Doppler gains and what removing the camera costs on the worst slots (supplementary
Sec.~S14 gives the tail intervals).

\begin{table}[t]\centering
\caption{Power-capture ratio: the received power a predictor's top-$1$ beam collects
relative to the optimal beam's, from the measured $64$-beam power vectors over
$774\times7$ test slots. The last two columns give the
tail: the share of slots losing more than $3$\,dB and the $95$th percentile of the loss.}
\label{tab:capture}\scriptsize
\setlength{\tabcolsep}{2pt}
\begin{tabular}{@{}lcccccc@{}}
\toprule
 & & \textbf{captured} & \multicolumn{4}{c}{\textbf{power lost}}\\
\cmidrule(lr){3-3}\cmidrule(lr){4-7}
\textbf{Predictor} & \textbf{ATop-1} & mean & med.\ [dB] & mean [dB] & $>\!3$\,dB [\%] & p95 [dB]\\
\midrule
RA only            & $37.3{\scriptstyle\pm1.2}$ & $0.856$ & $0.20$ & $0.91{\scriptstyle\pm0.09}$ & $8.2$ & $4.32$\\
DA only            & $41.8{\scriptstyle\pm1.6}$ & $0.887$ & $0.12$ & $0.67{\scriptstyle\pm0.09}$ & $4.8$ & $3.00$\\
RD only            & $39.8{\scriptstyle\pm1.1}$ & $0.877$ & $0.15$ & $0.74{\scriptstyle\pm0.08}$ & $5.8$ & $3.39$\\
RA$+$DA, no KD     & $41.3{\scriptstyle\pm0.9}$ & $0.890$ & $0.12$ & $0.65{\scriptstyle\pm0.05}$ & $4.7$ & $2.91$\\
RA$+$RD, no KD     & $41.9{\scriptstyle\pm1.0}$ & $0.893$ & $0.11$ & $0.64{\scriptstyle\pm0.08}$ & $4.2$ & $2.73$\\
RA$+$DA$+$RD       & $42.3{\scriptstyle\pm1.1}$ & $0.895$ & $0.11$ & $0.61{\scriptstyle\pm0.06}$ & $3.9$ & $2.58$\\
RA$+$RD, $+$KD     & $41.4{\scriptstyle\pm1.1}$ & $0.892$ & $0.12$ & $0.64{\scriptstyle\pm0.03}$ & $4.1$ & $2.61$\\
Vision-only        & $44.1{\scriptstyle\pm1.8}$ & $0.918$ & $0.08$ & $0.44{\scriptstyle\pm0.08}$ & $2.1$ & $1.98$\\
Teacher (V$+$R)    & $45.8{\scriptstyle\pm1.2}$ & $0.926$ & $0.05$ & $0.41{\scriptstyle\pm0.05}$ & $1.7$ & $1.73$\\
\midrule
\emph{spread} & $8.5$\,pp & & & $0.51$\,dB & $6.5$\,pp & $2.59$\,dB\\
\bottomrule
\end{tabular}
\\[2pt]
{\scriptsize Mean over ten seeds for every row; the median, the share of slots above
$3$\,dB and the $95$th percentile are of each seed's pooled slots, averaged over seeds.
The first seven rows are radar-only. The mean loss exceeds the median because a small
number of large misses dominate it; the share above $3$\,dB counts them against a
threshold relative to the best beam, not a measured outage. The teacher is
the Phase-2 Doppler--angle teacher of Table~\ref{tab:accuracy}. Differences quoted in the text are computed from unrounded means
and can differ from differences of the rounded entries by $0.01$\,dB or $0.1$\,pp.}
\end{table}

\needspace{5\baselineskip}
\section{The Design Choices in Received Power}
\label{sec:res_choices}
This section asks what explains the Doppler gain and the distillation null on the scale of
Sec.~\ref{sec:res_resolution}. Secs.~\ref{sec:res_ablation} and~\ref{sec:res_fusion} ask why the tested
distillation yields no measurable gain, and Sec.~\ref{sec:res_doppler} asks what in the
Doppler-resolved map supplies the measured gain. Fig.~\ref{fig:forest} collects the decibel
contrasts on which this section and
Sec.~\ref{sec:results} rest.

\subsection{Ablation: Distillation Signals}
\label{sec:res_ablation}
No tested distillation signal helps measurably. Adding the logit, feature and
temporal-attention terms cumulatively to the RA$+$RD student (Table~\ref{tab:accuracy})
changes received power by $-0.01$ to $-0.03$\,dB, every interval covering zero
(Fig.~\ref{fig:forest}; supplementary Sec.~S14 lists the three changes). We read these rows as no
measurable change, not a measured degradation. The null does not hinge on the teacher
checkpoint: in a pre-specified replication, the two other RA$+$RD teachers trained the
same way give $-0.01$ and $-0.03$\,dB (ten seeds each). The concatenation teacher of
Sec.~\ref{sec:res_fusion} gives the same null. This differs from same-modality KD, where
feature-level transfer helps~\cite{ma2025kd}, and is consistent with a modality gap: the
teacher's advantage is largely vision-specific.

The null holds over the tested losses and weights. The unit weights were not tuned, but the
logit term already dominates the objective early in training. The temporal-attention term has
little to carry: the teacher attends almost uniformly over the seven frames, and
students trained without the term are already within $0.07$ to $0.10$\,nats of it, which leaves the term little
to close. Varying the logit weight over two decades keeps the student within
$0.03$\,dB of no distillation (five seeds each; supplementary Table~S9). A two-layer projector
for the feature term leaves top-$1$ accuracy where the linear one does (supplementary
Sec.~S10). The distillation pipeline is therefore kept as the apparatus that measured the null, not as part
of the selected predictor.

\subsection{Alternative Explanations for the Null}
\label{sec:res_fusion}
None of the tested alternatives explains the null: not the teacher's fusion design and not a
larger teacher--student gap. First, the residual in~\eqref{eq:fused}
anchors the fused feature to the radar branch, and a modality-agnostic teacher might expose a
more transferable structure. We therefore trained a concatenation teacher,
$\mathbf{F}_{\mathrm{fused}}=\mathrm{MLP}([\mathbf{F}_V,\mathbf{F}_R])$, under an identical
protocol, split and seed set. At the task-loss stage, in its Doppler--angle version, it matches the cross-attention
teacher, and neither leads the vision-only student by two points, so most of the teacher's
lead over the reference student is vision's (supplementary
Table~S12).
Distilling from its RA$+$RD version into the same RA$+$RD student changes
received power by $+0.01$\,dB, indistinguishable from distilling the cross-attention teacher,
so the tested fusion topology does not explain the null.

Second, a larger teacher--student gap creates no measurable transfer route. The
range--angle-only student trails the RA$+$DA teacher's seed-$42$ checkpoint by
$8.3$\,pp, where the reference student trails its own source by $3.6$. Distilling that
checkpoint into the range--angle-only student changes received power by $+0.06$\,dB over ten seeds per arm, consistent
with the original null. These experiments do not determine whether the remaining
teacher information is radar-inexpressible or merely inaccessible to the tested losses.
Third, widening the undistilled student two- or fourfold changes received power by no more
than $0.04$\,dB (supplementary Sec.~S11); widened students were not distilled.

\begin{figure}[t]\centering
\includegraphics{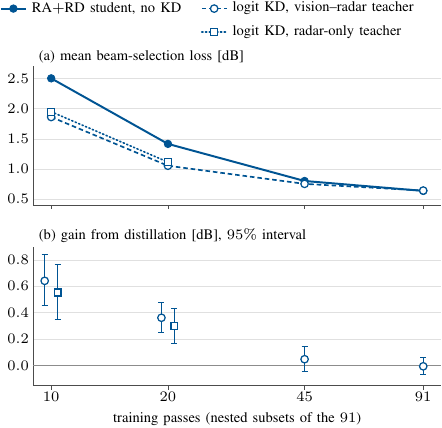}
\caption{Learning curve of the reference student (exploratory; five seeds per arm, ten at
$91$ passes), retrained on nested subsets of the $91$ training passes. (a) Mean
beam-selection loss without distillation, with logit distillation from the vision--radar
teacher and, at ten and twenty passes, from a radar-only teacher (the $91$-pass student).
(b) The gain from distillation with its $95\%$ interval. The gain shrinks as the student's
own data grow and is not measurable at full data; a radar-only teacher delivers most of it.}
\label{fig:lc}
\end{figure}

Distillation does help when the student lacks data, and a teacher without a camera
supplies most of that gain. Retrained on nested subsets of the $91$ training passes
(five seeds each, exploratory), the student gains $0.64$\,dB from distillation at ten passes,
and the gain falls monotonically to $-0.01$\,dB at all $91$ (Fig.~\ref{fig:lc}; supplementary
Table~S13). The null of
Sec.~\ref{sec:res_ablation} is therefore the data-rich end of a curve, not a sign that the
distillation pipeline fails. Because the teacher saw all $91$ passes, we also
distilled a radar-only teacher, the $91$-pass reference student, as an exploratory arm
with a decision rule written beforehand. It recovers $0.55$ of the $0.64$\,dB at ten passes; students of the vision--radar
teacher gain a further $0.09$\,dB on $[-0.09, +0.27]$, a difference left unresolved.

\needspace{5\baselineskip}
\subsection{What the Doppler-Resolved Map Supplies}
\label{sec:res_doppler}
The Doppler gain holds at every slot of the horizon. Over the horizon the optimal beam moves
by a mean of $10.5$ indices, about $15^\circ$, so long-term prediction requires
\emph{anticipating} that displacement. The Doppler-augmented student stays above the
range--angle baseline at every slot, every per-slot interval excluding zero, and its gain
grows from $0.235$\,dB at the current slot to $0.385$\,dB at $t\!+\!6$ (Fig.~\ref{fig:perslot}). The two ends
separate only marginally in decibels and not in top-$1$ accuracy, so the data are consistent with growth
but do not establish it. The gain also appears in
each half of the passes split at the median GPS speed or beam speed; these intervals establish
a gain in each half, not speed invariance. Supplementary Sec.~S12 tabulates the per-slot
and per-subgroup figures and splits the gain by optimal beam.

\emph{Velocity, or other Doppler content?} The controls below bound this question without
settling it. The tests do not support a plain angular-resolution reading. Nor does one velocity per angle
reproduce the gain for this encoder: a student reading each angle's Doppler centroid, with or
without its spread, is measurably worse than one reading the full distribution. Whether the
gain lies in the velocity ordering of the bins or in other properties of each distribution
remains unresolved.

\looseness=1 \emph{Which map structure contributes.} Retraining without range
localization in the second map causes no resolved loss of top-$1$ accuracy. On a locally straight road, range, angle and radial
velocity together fix the vehicle's velocity vector away from broadside, and hence its future angle, and a
range--angle map does not preserve the radial velocity. The controls of
Table~\ref{tab:controls} ask, one axis at a time, how much of that structure the encoder
uses. Permuting the range axis of the reference range--Doppler map destroys where in range the
moving returns sit while keeping each velocity bin's total. The student that reads those maps
has an interval against the unpermuted input that includes zero and one against range--angle
alone that excludes it, so destroying range localization costs at most $3.0$\,pp. The controls that act on the
velocity axis belong to the Doppler--angle map, whose rows are Doppler bins.

\emph{What pooling the Doppler axis costs.} In a separate sweep, about $32$ Doppler groups reach the
full-resolution mean for this encoder. We pooled the Doppler axis of the Doppler--angle map into $B$ groups, repeating each
group's mean back over its rows. The Doppler--angle-only student then trails its full-resolution
run by $5.5$\,pp at $B\!=\!1$ but by only $0.05$\,pp at $B\!=\!32$ (five seeds a point). The sweep's own baseline sits an unexplained $1.2$\,pp above Table~\ref{tab:accuracy}'s, so it is read only within itself (supplementary Sec.~S13). This concerns the
resolution the \emph{encoder} uses, not the resolution the radar must acquire.

Among single maps, the order follows whether the map carries Doppler, not whether it resolves
angle. The Doppler--angle map loses $0.67$\,dB, the angle-blind range--Doppler map $0.74$ and
the angle-resolving range--angle map $0.91$, so the angle-blind map leads the range--angle map by an indicative
$0.17$\,dB. Among the Doppler-resolved inputs a residual angle effect is visible but
unresolved. Doppler presence is therefore the first-order distinction across these inputs:
each Doppler-resolved one beats range--angle alone on an interval excluding zero.

\begin{table}[t]\centering
\caption{Controls on radar-map content: ATop-1 of each minus that of the arm it transforms
(Table~\ref{tab:accuracy} or the footnote). A permutation is drawn per
frame and applied to every column, so each column keeps its values but loses the axis
label; a mean collapses the axis and repeats it back, so the map keeps its shape. Supplementary
Table~S3 gives the status of each row.}
\label{tab:controls}\scriptsize
\setlength{\tabcolsep}{2.5pt}
\begin{tabular}{@{}lllc@{}}
\toprule
\textbf{Control} & \textbf{Destroys} & \textbf{Comparator} & \textbf{$\Delta$ATop-1 [pp]}\\
\midrule
\multicolumn{4}{@{}l@{}}{\emph{(a) axis transformations}}\\
DA rows permuted        & velocity label   & DA only  & $-1.53$ $[-3.73, +0.56]$\\
DA to per-angle mean    & the spread       & DA permuted & $-2.21$ $[-3.79, -0.61]$\\
DA binarized            & the magnitudes   & DA only  & $-2.23$ $[-4.71, +0.02]$\\
RD rows permuted$^{\dagger}$ & where in range & RA$+$RD & $-1.13$ $[-2.99, +0.66]$\\
RD to per-velocity mean & where in range & RD only & $+0.43$ $[-1.46, +2.32]$\\
\midrule
\multicolumn{4}{@{}l@{}}{\emph{(b) moment maps, with predictions written before training}}\\
DA total at centroid & spread, shape & DA only & $-4.59$ $[-7.18, -2.07]$\\
DA Gaussian moments & the shape & DA only & $-3.07$ $[-5.45, -0.68]$\\
DA Gaussian $+$ floor & the shape & DA only & $-2.97$ $[-5.32, -0.67]$\\
DA spread only & centroid, shape & DA only & $-3.99$ $[-6.46, -1.61]$\\
\bottomrule
\end{tabular}
\\[2pt]{\scriptsize $^{\dagger}$Five seeds against ten; every other row is ten against
ten. Transformed-arm ATop-1 (\%): DA permuted $40.2$, per-angle mean $38.0$,
binarized $39.5$, RD permuted $40.8$, RD per-velocity mean $40.2$, total at centroid
$37.2$, Gaussian $38.7$, Gaussian $+$ floor $38.8$, spread only $37.8$. Panel (b) replaces
each column's Doppler distribution by its total at its centroid row, by a Gaussian with its
centroid and standard deviation, by that Gaussian plus the column's median floor or by
that Gaussian centered at zero Doppler.}
\end{table}

The axis transformations of Table~\ref{tab:controls}(a) show that the encoder uses more than
each angle's mean over the Doppler axis. Collapsing each
column of the Doppler--angle map to its mean leaves the student indistinguishable from
range--angle alone, although the collapse keeps the moving target's angle profile. This is also
consistent with the shortfall of the classical readers of Sec.~\ref{sec:res_trivial}, which reduce the
profile to its peak. The same collapse along the range axis of the range--Doppler map causes
no measurable loss, so the effect depends on which axis is collapsed. Whether the velocity
\emph{ordering} matters we could not resolve: permuting the rows gives an interval that admits a
loss close to the whole Doppler gain, an unresolved null, not an established equivalence. Nor could we resolve whether the magnitudes inside the distribution matter: the
binary map sits between the full map and range--angle alone without separating from
either.

\emph{Which statistics of the distribution.} Four exploratory arms, with
predictions written before training, replace each column's distribution by its total at the
centroid, a Gaussian with its centroid and spread, that Gaussian plus the median floor, or
the spread alone (Table~\ref{tab:controls}(b)). None reaches the full map's accuracy: the column's total
placed at its centroid is $0.35$\,dB worse than the full map, and two of the
four predictions were wrong. The permuted map, which keeps each column's values but not their
order, sits above the collapse and the centroid arm but is separated neither from the
Gaussian nor from the full map (supplementary Sec.~S13). Each
arm also measures what this encoder reads from its format, not only the information the format
keeps.

\emph{Is the Doppler axis a finer angle sensor?} Not in the forms tested. On a one-directional straight road the
radial velocity is
$v\sin\vartheta$, so at near-constant speed the Doppler bin could act as a finer bearing sensor,
the reading Demirhan and Alkhateeb~\cite{demirhan2022radar} offer for their range--velocity
result. A test written down in advance (supplementary Sec.~S13) does not support its simplest form. A reader
mapping the Doppler centroid to a beam errs by $11.7$ indices RMS across passes, against
$6.8$ for the angle-axis peak of the same maps. Neither the pass speed nor its dispersion
repairs that error. The Doppler axis carries bearing information only jointly with quantities the map
does not fix. The geometric argument also delimits the cue: a moving-target map informs
future angle only while the target's direction persists, so we expect it to weaken through
turns and intersections.

\looseness=1 We also do not reproduce the ordering of~\cite{demirhan2022radar}. There, velocity
representations used as replacements for the range--angle map in single-slot prediction were
no better; here an angle-blind but Doppler-resolved map is better than range--angle alone, by an indicative margin.
That study used the same scenario and codebook. The two studies differ in the horizon (one slot against
seven), the input ($256$ range bins against $128$; their range--angle map sums per-chirp magnitudes, ours averages Doppler-bin magnitudes), the network and the split. Theirs is a
$70/30$ division of $6{,}319$ samples not described as respecting pass boundaries, which would
favor a position-memorizing range--angle map; ours is pass-chronological. The Doppler gain is
already present at the current slot (supplementary Table~S14), so the horizon alone does not explain the
difference, and we have no measurement isolating the rest. The gain is also visible in the
angular error itself, which needs no pattern beyond the codebook grid. On the $90^\circ$ field
of view, adding the range--Doppler map lowers the mean misalignment from $2.28^\circ$ to
$1.82^\circ$.

\section{Discussion and Conclusion}
\label{sec:conclusion}
\textbf{Main findings.} Within the tested family, the Doppler-resolved content a radar-only
predictor reads at inference adds measurable received power. The visual information only
its teacher sees in training, by contrast, does not measurably transfer through the tested
distillation routes with the full training set. A
range--Doppler map beside the
range--angle map gains $0.28$\,dB ($[0.14, 0.41]$), mostly in the tail, and an all-zero
second map does not. The gain replicates on Scenario~31 ($+0.14$\,dB as pre-specified, $+0.20$\,dB after a label correction),
where holding the last measured beam beats both radar arms. Distillation from a
motion-mask vision--radar teacher gains $-0.01$\,dB on the discovery site ($[-0.07, +0.06]$), an interval that
excludes recovering over a quarter of the source teacher's $0.24$\,dB advantage. The tested teacher, student,
fusion and distillation variants show the same null in the metric each was scored on.

\looseness=-1 \textbf{Design lessons.} First, the link-level size of an accuracy difference should be
measured, not assumed. Top-$1$ accuracy orders the predictors nearly as received
power does, but a point of it is worth only $0.062$\,dB of mean beam-selection loss here.
Second, the tested predictor benefits from a resolved Doppler axis. One velocity per angle, the
Doppler centroid with or without its spread, falls short of the full map for this encoder,
and the role of the velocity ordering is unresolved. Third, what a modality seen only in training adds should likewise be measured,
not assumed. Our null is a statement about the tested architectures and transfer routes, not
about cross-modal learning in general. It is consistent with a teacher whose advantage is
mostly vision's: a vision-only student reproduces four-fifths of it, a decomposition rather than an
isolated mechanism. Where distillation helps, for a ten-pass student (exploratory), a
radar-only teacher trained on all training passes reproduces $0.55$ of the $0.64$\,dB. The two teachers also
differ in architecture and size, so this does not isolate the camera, and teacher data and soft targets stay confounded. The radar-only pipeline removes the camera and $95\%$ of the parameters
but not a matching share of the arithmetic, and it does not improve accuracy.

\textbf{Scope.} Every learned contrast is conditional on one CNN--GRU family, though the
Doppler gain over range--angle alone survives separate input branches and a single repeated
frame (supplementary Sec.~S5). The family also needs data (Fig.~\ref{fig:lc}(a)): even $45$ passes leave the undistilled
student $0.16$\,dB ($[+0.07, +0.27]$) behind the $91$-pass arm. A site of this kind thus
appears to need on the order of a hundred labeled passes, at odds with the low-cost roadside
motivation. The principal estimates come from one site, vehicle and direction of travel at
pass-median speeds of $2.8$--$4.7$\,m/s ($5$th to $95$th percentile) in daylight and clear weather. The passes include no turns or intersections and do not enter the low-speed
regime in which the slow-time mean subtraction would remove the target with the clutter. The
windowed, open-loop protocol makes these results optimistic benchmarks, not deployed
performance. The labels come from exhaustive sweeps that a deployed link would pay for, and
each window is aligned to the frame clock rather than to the link's acquisition timing. Both
conditions are the same for every arm, so the comparisons are like for like. The slope of
Sec.~\ref{sec:res_resolution} and the agreement in ordering are local to this predictor set,
codebook and site. A different codebook, angular spread or predictor error pattern can change
both, so applying the slope elsewhere requires new measured sweeps.

\textbf{Limitations and open questions.} The data cannot say whether the encoder reads the
sign of the Doppler shift or only its magnitude, and the one same-road day/night pair cannot
isolate illumination (Sec.~\ref{sec:res_pred}). Every model learns the arg\,max beam through
a classification loss (Sec.~\ref{sec:problem}). Whether training on the measured power
vectors, which aims more directly at received power, would change either finding is
untested. Soft labels from a teacher trained on the same ten passes
(Sec.~\ref{sec:res_fusion}) were not tested either. The power-capture ratio is per slot, without
scheduling or retransmission. Each Scenario~9 slot's loss is capped by its sweep's dynamic range
($9.7$\,dB at the median), which bounds the tail statistics in scale but not the ordering of
the arms (supplementary Sec.~S9). The radar-only case is argued against a camera, not against
a position report from the vehicle~\cite{charan2022vision}. The road-coordinate tracker is the strongest classical baseline we could build
without a detection and association pipeline, not a bound on classical processing. The
latency figures are model-side inference times from~\cite{companion_voi}; capture,
preprocessing and run-time memory were not measured. Every result rests on measured narrow-beam
sweeps, taken as simultaneous within a sample and as valid in both link directions (beam
reciprocity), neither of which the dataset documents. Beamwidth adaptation, as
in~\cite{orimogunje2026beamguard} on the same dataset, is not analyzed. Whether radar degrades
more gracefully than vision, and whether moving-target cues transfer better to unseen
environments, are natural next steps.

\raggedbottom
\section*{Acknowledgment}
The authors used Claude (Anthropic) to assist with editing the text (e.g., grammar) and with preparing figures and code. The authors reviewed and verified all such content and take full responsibility for it.

\bibliographystyle{IEEEtran}
\bibliography{library}
\end{document}